\documentclass[prl,aps,amssymb,twocolumn,reprint,superscriptaddress]{revtex4-1}
\usepackage{xeCJK}
\usepackage{amsmath}
\usepackage{amssymb}
\usepackage{amsthm}
\usepackage{amsfonts}
\usepackage{listings}
\usepackage{enumerate}
\usepackage{latexsym}
\usepackage{color}
\usepackage{xcolor}
\usepackage{bm}
\usepackage{hyperref}
\hypersetup{
 pdfnewwindow=true, colorlinks=true,
 linkcolor=blue, anchorcolor=blue,
 citecolor=blue, filecolor=blue,
 menucolor=blue, urlcolor=blue}

\usepackage{psfrag}

\usepackage{graphicx}
\usepackage{subfigure}
\usepackage{lipsum}
\usepackage{float}
\usepackage{array}
\usepackage{soul} 
\usepackage{titlesec}

\newcommand{\beginsupplement}{%
        \setcounter{table}{0}
        \renewcommand{\thetable}{S\arabic{table}}%
        \setcounter{figure}{0}
        \renewcommand{\thefigure}{S\arabic{figure}}%
        \setcounter{equation}{0}
        \renewcommand{\theequation}{S\arabic{equation}}
     }
     
\begin{document}
\soulregister{\cite}7 
\soulregister{\ref}7
\tolerance 10000

\draft

\title{Coexistence of topologically nontrivial and trivial insulating states in topological Anderson Chern insulator}


\author{Bo Yin}
\affiliation{Beijing National Laboratory for Condensed Matter Physics,
and Institute of Physics, Chinese Academy of Sciences, Beijing 100190, China}
\affiliation{University of Chinese Academy of Sciences, Beijing 100049, China}

\author{Yan Zhang}
\affiliation{Beijing National Laboratory for Condensed Matter Physics,
and Institute of Physics, Chinese Academy of Sciences, Beijing 100190, China}
\affiliation{University of Chinese Academy of Sciences, Beijing 100049, China}

\author{Anqi Wang}
\affiliation{Beijing National Laboratory for Condensed Matter Physics,
and Institute of Physics, Chinese Academy of Sciences, Beijing 100190, China}
\affiliation{University of Chinese Academy of Sciences, Beijing 100049, China}

\author{Jie Shen}
\affiliation{Beijing National Laboratory for Condensed Matter Physics,
and Institute of Physics, Chinese Academy of Sciences, Beijing 100190, China}
\affiliation{University of Chinese Academy of Sciences, Beijing 100049, China}

\author{Zhijun Wang}
\affiliation{Beijing National Laboratory for Condensed Matter Physics,
and Institute of Physics, Chinese Academy of Sciences, Beijing 100190, China}
\affiliation{University of Chinese Academy of Sciences, Beijing 100049, China}

\author{Quansheng Wu}
\email{quansheng.wu@iphy.ac.cn}
\affiliation{Beijing National Laboratory for Condensed Matter Physics,
and Institute of Physics, Chinese Academy of Sciences, Beijing 100190, China}
\affiliation{University of Chinese Academy of Sciences, Beijing 100049, China}

\begin{abstract}
 The interplay between disorder and topology has become a central theme in condensed matter physics. Disorder can not only destroy topological phases but also induce them, as exemplified by the topological Anderson insulator (TAI). Here we show that, in close analogy, disorder can drive the clean-limit, time-reversal-broken(T-broken) quantum spin Hall state of ferromagnetic(FM) monolayer $\mathrm{MnBi_4Te_7}$ into a quantum anomalous Hall phase, which was called topological Anderson Chern insulator (TACI). Using density functional theory (DFT) and nonequilibrium Green’s function (NEGF) calculations in the presence of disorder, we identify disorder induced phases—including T-broken TAI, TACI, Normal insulator, etc., then construct a comprehensive phase diagram. To discriminate multiple phases in the strong disorder regime, we further use the density of states computed within the self-consistent Born approximation (SCBA), which in particular distinguishes gapped and ungapped topological phases. We find that the two effective band inversions of Hamiltonian are suppressed at distinct critical disorder strengths; the survival of a single inversion over a finite disorder window stabilizes the TACI. The region of quantized Hall resistance is not confined to the band gap but extends into a mobility gap, substantially enlarging the parameter space exhibiting quantization. Remarkably, at strong disorder, we further propose a zero Hall plateau insulating state characterized by an insulating bulk and edge channels subject to diffusive scattering that can coexist with the TACI. This behavior is distinct from a conventional band-gap Chern insulator and provides a clear experimental signature. Compared with the clean region, the expansion of the energy window supporting the Chern insulator of the disordered region enhances thermal robustness and enables observation of the quantum anomalous Hall effect at higher, technologically relevant temperatures, advancing both fundamental studies and potential applications.

\end{abstract}
\date{2026.5.21} 
\maketitle

In recent years, topological states of matter have become one of the most active frontiers in condensed matter physics. Among them, the quantum anomalous Hall effect (QAHE) \cite{PhysRevLett.90.206601,PhysRevLett.106.166802,PhysRevLett.61.2015,RevModPhys.95.011002} holds a particularly prominent place. It realizes one dimensional, dissipationless transport via chiral edge states in topological materials without an external magnetic field, with profound implications for low-power electronics and topological quantum computation.

Experimentally, the QAHE can be achieved either by magnetically doping topological insulators—e.g., introducing Cr or V into three dimensional topological insulators such as $\mathrm{(Bi,Sb)}_{2}\mathrm{Te}_{3}$ \cite{doi:10.1126/science.1187485,doi:10.1126/science.1234414,Mogi2015APL} or by employing intrinsic magnetic topological materials such as $\mathrm{MnBi}_{2}\mathrm{Te}_{4}$, the first intrinsic magnetic topological insulator \cite{PhysRevLett.122.206401,doi:10.1126/sciadv.aaw5685,YanGong:76801,doi:10.1126/science.aax8156,Otrokov2019AFMTI}, which has stimulated extensive subsequent studies \cite{PhysRevX.9.041040,10.1093/nsr/nwaa089,Liu2020_MnBi2Te4,Liu2020AxionAFMTI,Zhao2021_MnBi2Te4,Wang2025AlOxQAHE,Lian2025AFM_QAHE}. QAHE is also observed in other systems, such as moiré materials with 
orbital ferromagnetism\cite{doi:10.1126/science.aay5533,Chen2020,Li2021}, and spin-orbit proximitized graphene\cite{doi:10.1126/science.adj8272,doi:10.1126/science.adk9749}.

To broaden the pool of QAHE candidate systems, attention has turned to the homologous series $\mathrm{MnBi}_{2n}\mathrm{Te}_{3n+1}$ \cite{npj,ALIEV2019443}. Members of this family are built from septuple layer (SL) blocks ($\mathrm{MnBi}_{2}\mathrm{Te}_{4}$) and quintuple layer (QL) blocks ($\mathrm{Bi}_{2}\mathrm{Te}_{3}$), with $n$ QLs inserted between adjacent SLs. The $n=1$ member, $\mathrm{MnBi}_{4}\mathrm{Te}_{7}$, has accordingly drawn widespread interest \cite{PhysRevX.10.031013,PhysRevX.9.041065,PhysRevMaterials.6.054203}. Previous work \cite{PhysRevLett.123.096401} revealed that monolayer $\mathrm{MnBi_4Te_7}$ hosts a time-reversal–broken quantum spin Hall state\cite{PhysRevLett.107.066602}, however, a recent experiment\cite{wang2025observationtopologicalandersonchern} has demonstrated that the QAHE can also be realized in this system.

One of the main challenges for QAHE is the typically small bulk energy gap which makes the topological state fragile and restricts its operation to ultra-low temperatures, often below one kelvin in current experimental realizations\cite{RevModPhys.95.011002}. Such stringent temperature requirements severely limit the potential for device applications and hinder the observation of QAHE under more practical conditions. Therefore, enlarging the energy gap and improving the stability of QAHE materials remain critical goals for advancing both fundamental research and technological applications.
\begin{figure*}[ht]
\centering
\includegraphics[width=1.0\textwidth]{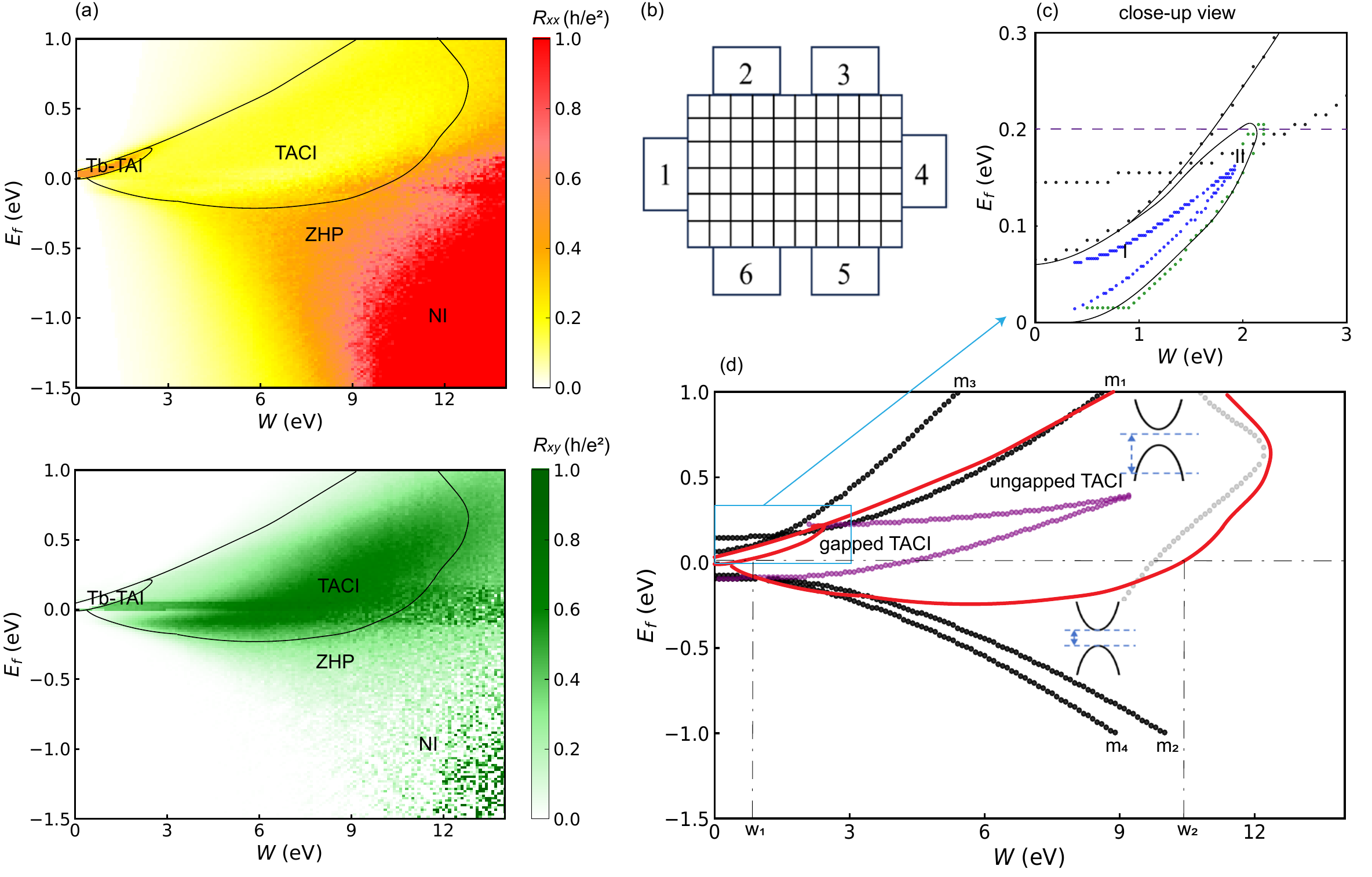}
\caption{Transport calculations by NEGF. (a)Phase diagram calculated by NEGF.  $R_{xx}, R_{xy}$ versus disorder strength $W$ and Fermi energy $E_f$. The size is chosen as 200a$\times$200a, a is the lattice constant, with 200 averages. $R_{xx}$ and $R_{xy}$ are plotted in red and green color boxes with phase boundaries in black lines, respectively.(b)Schematic diagram of six-terminal Hall bar. (c)Zoom-in on the light blue lines from Fig. 1(d). Region I, demarcated by the blue line, represents gapped TAI($\rho_2=0$). Region II, demarcated by the green line, represents ungapped TAI($\rho_2 \le$ 1/$\Delta_2$). (d)Phase diagram with phase boundaries only(red lines). Black lines are effective band edges by SCBA labeled as $m_1$ to $m_4$ in which $m_1$ and $m_2$ denote upper-left block of Eq. (S1), $m_3$ and $m_4$ for the other. Purple and gray lines are contour lines of density of states for upper-left block of Eq. (S1) by SCBA. In the region enclosed by the purple line, $\rho_1$=0, the gray line indicates $\rho_1$=1/$\Delta_1$. The dash–dotted lines correspond to $E_f=0.04$ eV, $W=w_1$ and $w_2$ mark the transition points between TAI, TACI, and NI.  }\label{fig:negf}
\end{figure*}

QAHE is a manifestation of a nontrivial topological phase. Topological phases are characterized by robustness to weak disorder and imperfections, protected by global invariants that remain unchanged under continuous perturbations; yet sufficiently strong disorder can destroy this protection. The interplay between topology and disorder has therefore attracted broad attention \cite{PhysRevLett.98.256801,Ryu_2010,PhysRevLett.109.246605,PhysRevB.91.235111}. Importantly, disorder does not invariably destroy topology; it can also induce it. A canonical example is the topological Anderson insulator (TAI) \cite{PhysRevLett.103.196805,PhysRevB.80.165316,PhysRevLett.102.136806,PhysRevLett.105.216601,PhysRevLett.105.216601}. Prior work has shown that the TAI regime can extend beyond the band gap into a mobility gap and persist over a broad range of disorder strengths independent of disorder type\cite{PhysRevB.85.195125,PhysRevB.85.035107}. Experimentally, TAI phenomena have been realized in platforms including cold atoms \cite{doi:10.1126/science.aat3406}, photonic crystals \cite{PhysRevLett.125.133603,PhysRevLett.132.066602}, electric circuits \cite{PhysRevB.100.184202}, and HgTe-based semimetals \cite{PhysRevResearch.7.L022033}.

It is widely recognized that QAHE arises from a topological transition driven by exchange coupling that breaks time-reversal symmetry (TRS)\cite{PhysRevLett.101.146802,doi:10.1126/science.1187485}. Inspired by the TAI mechanism, an alternative route to QAHE becomes possible: even in the absence of a preexisting, magnetically induced band inversion, disorder alone can trigger the topological phase\cite{PhysRevB.93.214206,PhysRevB.102.201405,chen2021evolution}. 

Here, we investigate disorder-induced topological phases in ferromagnetic(FM) monolayer ${\rm MnBi_4Te_7}$. We incorporate realistic material parameters into nonequilibrium Green’s-function quantum transport simulations with Anderson disorder to demonstrate the existence of a topological Anderson Chern insulator (TACI) \cite{wang2025observationtopologicalandersonchern,li2024reentrantquantumanomaloushall}, and establish the complete phase diagram. To elucidate the phase structure across the diagram, we introduce a density-of-states criterion from the self-consistent Born approximation to distinguish phases, with particular efficacy in the strong-disorder regime where it cleanly separates gapped from ungapped regions. Local current distributions and edge
projection of disordered energy spectrum are further used to identify nonlocal states.

\textbf{Transport calculations by the nonequilibrium Green’s function method(NEGF).} First-principles calculations are employed to examine the electronic structure of the ferromagnetic (FM) monolayer ${\rm MnBi_4Te_7}$ to demonstrate it is a time-reversal-broken(T-broken) quantum spin Hall state. Starting from quantum spin Hall\cite{PhysRevLett.95.146802,Bernevig2006ScienceQSH}, when TRS is weakly broken by a local perturbation that does not close the bulk gap, the spin Chern number remains unchanged. The Kramers degeneracy is lifted, so Kramers protection is lost, however, the existence of a pair of gap traversing edge modes is retained as long as the bulk gap stays open and the spin Chern number remains quantized, which was called T-broken quantum spin Hall state. Details of the first-principles methods, computational parameters and model are provided in the supplemental material \cite{sup}. 

To simulate Hall responses and disorder effects, We calculate transport properties of the six-terminal Hall-bar device as in Fig.\ref{fig:negf} (b) by the Landauer-Büttiker formula \cite{Landauer1970,Buttiker1988,FisherLee1981} and NEGF \cite{PhysRevB.50.5528,PhysRevLett.68.2512,datta}. The current is applied between lead
1 and lead 4, the voltage is measured between lead 2 and
lead 3 to get the longitudinal resistance $R_{xx}$ and lead 2 and
lead 6 to get the Hall
resistance $R_{xy}$. The current in the lead p with spin index $\sigma$ is \cite{datta}

\begin{gather}
I_{p \sigma}=\frac{e}{\hbar} \sum_{q \neq p} T_{p q}^\sigma\left(V_{p \sigma}-V_{q \sigma}\right),
\end{gather}

where $T_{p q}^\sigma$ is the transmission coefficient from lead p to lead q with spin $\sigma$, and $V_{p \sigma}$ stands for the voltage in the lead p with spin index $\sigma$.
The transmission coefficient can be obtained from Green's function\cite{datta}

\begin{gather}
T_{\mathrm{pq}}=\operatorname{Tr}\left[\Gamma_{\mathrm{p}} \mathrm{G}^{\mathrm{r}} \Gamma_{\mathrm{q}} \mathrm{G}^{\mathrm{a}}\right],
\end{gather}

where $\Gamma_{p \sigma}=i\left[\Sigma_{p \sigma}^{\mathrm{r}}-\Sigma_{p \sigma}^{\mathrm{a}}\right]$ is the linewidth function, $\Sigma_{p \sigma}^{\mathrm{r}}$ is the retarded self-energy at the lead p with spin $\sigma$. For simplicity, the leads are chosen to have the same properties as the central region, but in the clean limit, in order to calculate the retarded self-energy with surface Green’s function\cite{PhysRevB.102.201405}.

The retarded Green’s function of the system can be calculated from\cite{datta}
\begin{gather}
G^r=\left[G^a\right]^{\dagger}=\left(E_f I-H_c-\sum_{p \sigma} \Sigma_{p \sigma}^r\right)^{-1},
\end{gather}
where $E_f$ is the Fermi energy and $H_c$ is the Hamiltonian of the central region.

In practice, we choose square central region to reduce edge-state scattering to the maximum extent, the size of lead is 40\% of the central length, leaving 30\% on the both sides of lead 1 and lead 4. 

Next, we consider the on-site random magnetic Anderson disorder $H_D=\sum_{i}\left\langle i| V_i |i\right\rangle$, where $\left|i\right\rangle$ is the basis including orbital and spin degrees of freedom, and each $V_i$ is uniformly distributed in the range $[-\frac{W}{2},\frac{W}{2}]$ with the disorder strength $W$ in units of eV. These simulations account for potential impurities, vacancies, and other forms of disorder likely to arise in real materials. Due to computational resource limitations, only small system sizes(N$\le$1000) are tractable at moderate computational cost. To address this, a finite-size scaling analysis is employed, where simulations are performed on a series of samples with varying system sizes, denoted as N.

Fig.\ref{fig:negf} (a) display the phase diagram of $R_{xx}$ and $R_{xy}$ with respect to disorder strength W and chemical potential $E_f$, and phase boudaries are plotted. Owing to the small band gap, the calculated $R_{xx}$ in the QAHE exhibits a small but finite value\cite{Chen2023ChiralEdgeCouplingQAHE}. Fig.\ref{fig:negf} (d) presents a phase diagram in which only the phase boundaries are indicated. The red lines correspond to the boundaries in Fig.\ref{fig:negf} (a) and others delineate a finer subdivision of the phases. Fig.\ref{fig:negf} (c) is a zoom-in of the region indicated by the light blue dashed lines, highlighting the T-broken TAI phases. In the clean limit $W$=0, $R_{xx} = 0.5 h/e^2$ in the band gap corresponding to the T-broken QSHE state. In the presence of disorder, three phases emerge: T-broken TAI, TACI, and Normal insulator(NI). Without loss of generality, we discuss the phase transition at $E_f=0.04$ eV indicated the dot-dash line in Fig.\ref{fig:negf} (d). As disorder strength increases from 0 to $w_1$, a the T-broken QSHE evolves into the T-broken TAI phase, with $R_{xx} = 0.5h/e^2$ corrsponding to orange region in Fig.\ref{fig:negf} (a), resembling the phase diagram of BHZ model from earlier work\cite{PhysRevLett.102.136806}. As disorder strength increases from $w_1$ to $w_2$, the T-broken TAI phase disappears while a TACI phase emerges, whose Fermi-energy window is substantially larger than the clean band gap \cite{PhysRevB.103.075434}.  With additional increase of $W$, the TACI is destroyed by Anderson transition\cite{PhysRevB.104.195416}, and the system ultimately enters the Anderson insulator(AI) in the thermodynamic limit \cite{PhysRev.109.1492,PhysRevB.92.085410}.
\begin{figure}[htb]
  \centering
  \includegraphics[width=\columnwidth]{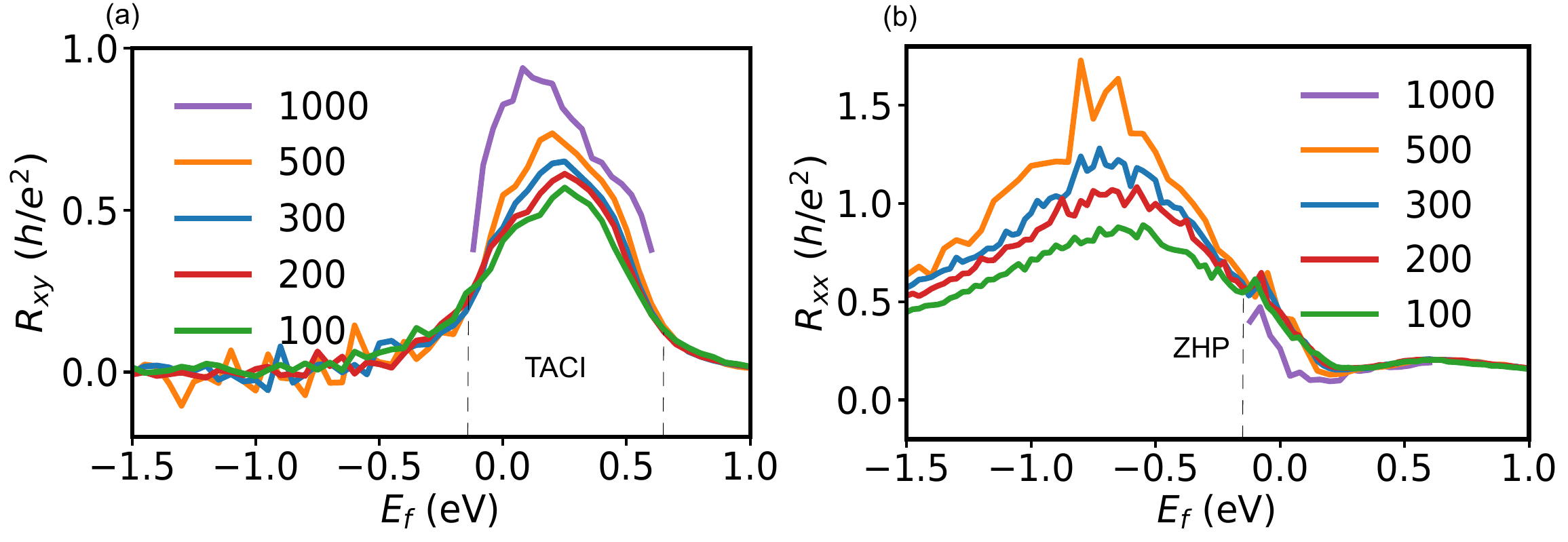}
\caption{Transport calculations by NEGF. (a) $R_{xy}$ versus Fermi energy $E_f$ at $W$ = 8.5 eV.(b) $R_{xx}$ versus Fermi energy $E_f$ at $W$ = 8.5 eV. We averaged over 2000 samples for system sizes N=100, 200 and 300, over 400 samples for size 500, and over 100 samples for size N=1000.}\label{fig:85}
\end{figure}
\begin{figure*}[ht]
\centering
\includegraphics[width=1.0\textwidth]{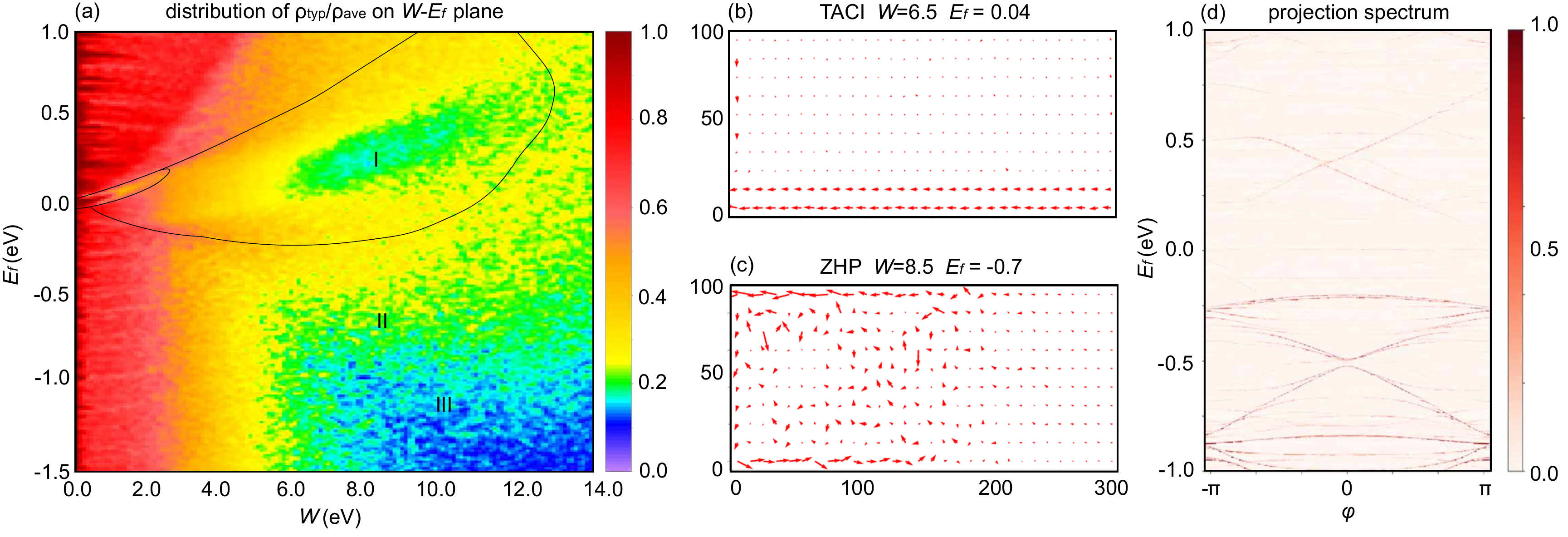}
\caption{Numerical evidence supporting edge states. (a)$\rho_{\text {typ}}$/$\rho_{\text {ave }}$ is calculated in a square sample with size 100 under periodic boundary conditions on the W-$E_f$ plane. (b)Current distribution for device of size 300*100 at $W$=6.5 eV, $E_f$=0.04 eV. (c)Current distribution for device of size 300*100 at $W$=8.5 eV, $E_f$=-0.7 eV. The calculations are averaged over 5000 configurations. (d)Edge projection of disordered energy spectrum of generalized momentum space, device of size 400*150 at $W$=8.5 eV.} \label{fig:nonlocal}
\end{figure*}

Specifically, we examine the disorder regime W=8.5 eV, the existence of the TACI phase is supported by a finite size scaling analysis. Fig.\ref{fig:85} (a) and (b) show the variations of $R_{xy}$ and $R_{xx}$ for different sizes. Here we notice that there are two crossing points of $R_{xx}$ and $R_{xy}$ curves at $E_f\sim$ -0.1 eV and 0.6 eV which are labeled as dashed lines. For $E_f<-0.1$ eV, $R_{xx}$ increases as the system size increases from 100 to 500, while $R_{xy}$ vanishes, indicating a trivial insulating phase with zero Chern number, similar features has been reported\cite{PhysRevLett.117.126802}. We refer to this state as zero Hall plateau(ZHP) insulating state, analogous phases have been discussed in both theory\cite{PhysRevB.89.205431,PhysRevLett.122.026601} and experiment\cite{Liu2021ZeroHallPlateauMnBi2Te4}, the details are discussed later. For $E_f>-0.1$ eV, $R_{xy}$ increases as the system size increases from 100 to 500, approaching a nearly quantized value at a size of 1000, while $R_{xx}$ vanishes, suggesting a Chern insulator with no clean-limit counterpart. The results obtained from the transport calculations are in close agreement with experimental observations \cite{wang2025observationtopologicalandersonchern}. The $R_{xy}$ curves for various system sizes converge at two crossing points, marking the phase transition points in the thermodynamic limit, clear numerical signatures of phase boundaries have likewise been reported in earlier studies of the TAI \cite{PhysRevB.109.L201102}.

\textbf{The self consistent Born approximation method (SCBA).} In the previous study, the disorder-induced topological phase transitions of TAI can be
 understood  with the help of an effective medium theory and the SCBA method\cite{PhysRevLett.103.196805,PhysRevLett.115.246603,PhysRevLett.125.166801}. However, conventional methods break down in the strong-disorder regime, TAI-I and TAI-II are introduced to differentiate gapped and ungapped bulk states\cite{PhysRevB.85.195140,PhysRevLett.132.066602}. To understand the phase diagram, we emphasize that disorder-induced density of states $\rho$ inside the effective bulk gap play a critical role, and provide a natural criterion for delineating the phase boundaries among these phases within the SCBA framework. $\rho$ is defined as \cite{PhysRevB.106.134201}
 
 \begin{gather}
\rho\left(E_f,W\right)=-\frac{1}{\pi}\left[\operatorname{Tr} \int \frac{d^2 \boldsymbol{k}}{E_f-H(k)-\Sigma}\right],
\end{gather}
where $\Sigma$ is the self energy induced by the disorder obtained from SCBA.

The phase boundaries predicted by SCBA effective theory are indicated as black, blue, purple, green and gray dotted lines in Fig.\ref{fig:negf} (c) and (d). In the schematic inset of Fig.\ref{fig:negf} (d), the blue dashed lines denote disorder-induced energy levels, and the dashed arrows indicate their mean level spacing. In the clean limit no states reside inside the bulk gap, i.e., $\rho=0$. With increasing disorder, the SCBA-predicted evolution of the effective band edges(black lines of $m_1$ to $m_4$) accurately captures the topological transition in the weak-disorder regime, details are shown in the Supplemental Material\cite{sup}. Within the effective band-inverted region(in between two pairs of dotted black lines in (c)), $\rho=0$ signals a gapped phase where band structures are renormalized by disorder; as disorder gradually introduces in-gap states and $\rho>0$ (blue and purple dotted lines in (c) and (d)), the spectral gap closes and a transition from gapped to ungapped behavior occurs. Crucially, in-gap impurity states alone do not alter the topological character: edge states persist while coexisting with impurity states. When the disorder-induced DOS satisfies $\rho\sim1/\Delta$ (green and gray dotted lines in (c) and (d)), the average impurity-level spacing becomes comparable to the effective bulk gap $\Delta$. Counter-propagating edge channels then couple via impurity states at nearby energies, enabling Anderson transitions.

\textbf{Further analysis of nonlocal states.} To further clarify whether the phases exhibit extended or localized behavior, $\rho_{\text {ave}}$ and $\rho_{\text {typ}}$ are introduced by summing over local density of states $\rho\left(i, E_F\right)$ by exact diagonalization\cite{RevModPhys.78.275,PhysRevB.88.195145,PhysRevLett.127.236402}
$$
\begin{aligned}
\rho\left(i, E_f\right) & =\sum_{n, \alpha}|\langle i, \alpha \mid n \rangle|^2 \delta\left(E_f-E_{n}\right) \\
\rho_{\mathrm{ave}}\left(E_f\right) & =\left\langle\left\langle\rho\left(i, E_f\right)\right\rangle\right\rangle \\
\rho_{\mathrm{typ}}\left(E_f\right) & =\exp \left[\left\langle\left\langle\ln \rho\left(i, E_f\right)\right\rangle\right\rangle\right] 
\end{aligned}
$$
where $\langle\langle\ldots\rangle\rangle$ is the arithmetic average over the sample sites and disorder realizations, $|i, \alpha\rangle$ denotes an eigenstate at site $i$ and orbital $\alpha$, and $n$ is the index for energy level. For extended states, $\rho_{\text{typ}}/\rho_{\text{ave}}\to \text{finite}$. For localized states, $\rho\left(i, E_f\right)$ vanishes on certain sites $i$, causing $\rho_{\text{typ}}$ to approach zero. Therefore, different values of $\rho_{\text{typ}}/\rho_{\text{ave}}$ reflect distinct localization characteristics.

In Fig.\ref{fig:nonlocal} (a), a green region I with a relatively small value of $\rho_{\text{typ}}/\rho_{\text{ave}}\sim0.2$ appears, and it corresponds to the TACI phase, indicating a localized state. Notably, in the green region II and the blue region III, $\rho_{\text{typ}}/\rho_{\text{ave}}$ is also very small. According to the transport calculations above, these regions correspond to a trivial insulating state; however, regions II and III exhibit different $\rho_{\text{typ}}/\rho_{\text{ave}}$ values, approximately 0.2 and 0.1, respectively, implying distinct localization behaviors in finite sizes. To further elucidate the nature of these regions, we next examine the edge-state transport properties.

To visualize the edge transport, we compute the nonequilibrium local current distribution within the NEGF framework. The local current between the neighboring sites i and j is\cite{PhysRevB.80.165316,PhysRevB.93.214206}
\begin{gather}
J_{\mathbf{i} \rightarrow \mathbf{j}}=\frac{2 e^2}{h} \operatorname{Im}\left[\sum_{\alpha, \beta} H_{\mathbf{i} \alpha, \mathbf{j} \beta} G_{\mathbf{j} \beta, \mathbf{i} \alpha}^n\left(E_f\right)\right]\left(V_L-V_R\right)
\end{gather}

where $V_L,V_R$ are the voltages at Lead L,R and are regraded as constant for all sites. 
 $G^n=G^r \Gamma_L G^a$ is electron correlation function with $G^r,\Gamma_L$ referring to Eq.(2). For site (i, j), current $J_x$ is calculated by summing all the local current of spins and orbitals from (i, j) to (i+1, j), while $J_y$ for (i, j) to (i, j+1). The total current for site (i, j) is a vector $\vec{J} = J_x \hat{e}_x + J_y \hat{e}_y$. 

The calculated results are shown in Fig.\ref{fig:nonlocal} (b) and (c). In the TACI regime, bond-current maps reveal chiral flow confined to the bottom edge, while the top edge is suppressed because its propagation direction is opposite to the applied bias. In the II regime of Fig.\ref{fig:nonlocal} (a), edge-like channels appear on both sides, as in the TAI, but quantization is lost due to diffusive backscattering of the edge modes, which is called ZHP state. Correspondingly, the transmission to the adjacent lead, evaluated from Eq. (2), falls below unity. In the thermodynamic limit, this phase flows toward NI; thus the apparent edge channels are a mesoscopic effect expected only in finite-size numerics and micron-scale experimental devices. Together with the SCBA results in Fig.~\ref{fig:negf}(d), it can be seen that the ZHP state lies within the effective band-inverted region; however, the excessively large disorder-induced DOS destroys the edge states, giving rise to the ZHP phase. In region III of Fig.\ref{fig:nonlocal}(a), edge states are nearly absent, indicating that it is a NI.

To further substantiate the edge character, we project the disordered spectrum onto generalized momentum space \cite{PhysRevB.82.115122,Chen_2012,PhysRevB.85.035107}. The relevant energy window is densely populated by impurity-induced states: band gap becomes ill-defined while a mobility gap emerges. Distinct, non-degenerate Dirac cones are resolved in different regimes—a single cone in the upper (TACI) regime and cones in the lower (ZHP) regime. The evolution of these cones with disorder, together with computational details, are provided in the Supplemental Material \cite{sup}.
\begin{figure}[htb]
  \centering
  \includegraphics[width=\columnwidth]{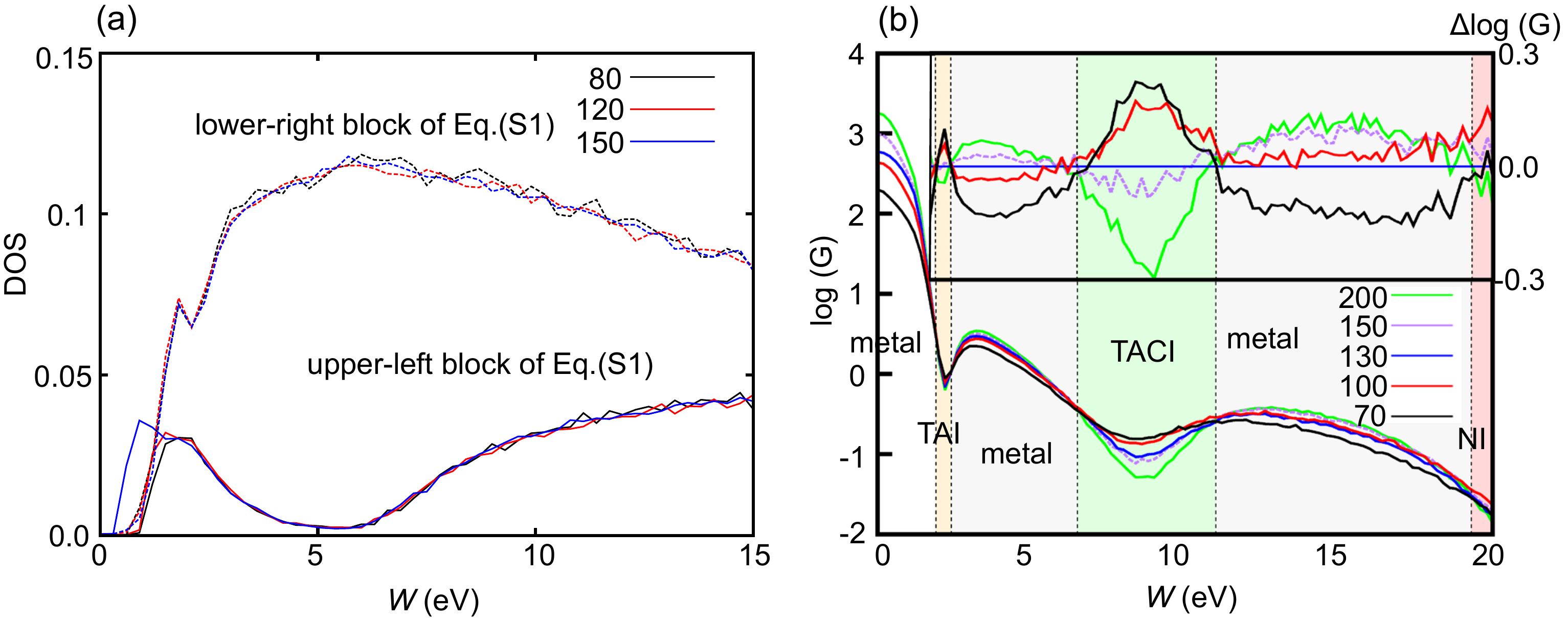}
\caption{Scaling calculations of phase transitions. (a)The dependence of the averaged density of states per site on disorder strength with size N=80,120,150 at $E_f$=0.2 eV under periodic boundary conditions. Solid and dashed lines denote upper-left and lower-right blocks of Eq. (S1), respectively. (b)Scaling of logarithmic average of the bulk conductance with size N=70,100,130,150,200 at $E_f$=0.2 eV under periodic boundary conditions. The upper inset is the difference relative to 130. The calculations were performed by averaging over 4000 realizations.} 
\label{fig:iqhe}
\end{figure}

\textbf{Scaling behavior in multiple phase transitions.} Fig.\ref{fig:iqhe} (a) presents the local density of states (LDOS)\cite{PhysRevB.88.045429,PhysRevB.85.195140} for different size of systems N=80,120,150 at $E_f$=0.2 eV along the purple dashed line indicated in Fig.\ref{fig:negf} (c), and the curves coincide indicating that finite-size effects can be neglected. Solid and dashed lines denote LDOS of upper-left and lower-right blocks of Eq. (S1), respectively. We find a pronounced dip at the bottom, which corresponds to the edge-transport signature as TAI pointed out in \cite{PhysRevB.85.195140}, indicating that the TACI originates from a single Hamiltonian block.

In disordered systems, delocalized states are a hallmark of disorder-driven topological phase transitions \cite{PhysRevLett.98.076802,PhysRevLett.105.115501}. To identify such states in the thermodynamic limit, we evaluate the logarithmic of the bulk conductance log(G) \cite{Chen_2012,PhysRevLett.86.3594,PhysRevB.88.014201}, for multiple system sizes N=70,100,130,150,200 at $E_f$=0.2 eV along the purple dashed line indicated in Fig.\ref{fig:negf} (c), with results shown in Fig.\ref{fig:iqhe} (b). As the disorder strength increases, the system undergoes a sequence of transitions, metal→T-broken TAI→TACI→NI, with extended states emerging in the vicinity of each phase boundary. In prior work, the delocalized state separating TAI and NI was attributed to a competition between the bulk localization length and the sample width \cite{Chen_2012}, which enables coupling between counter-propagating edge channels across the sample and the formation of a percolation state\cite{PhysRevB.88.014201,PhysRevB.91.214204}.
Our system explicitly violates time-reversal symmetry and thus belongs to the 2D unitary class (class A) \cite{PhysRevB.55.1142,PhysRevB.78.195125}. Typically, electronic states in this class are localized for any finite disorder, with delocalization occurring only at quantum Hall plateau transitions \cite{Wei1988}.
An analogous mechanism underlies the delocalized states at the T-broken TAI–TACI and TACI–NI boundaries, mirroring the extended states that appear at plateau–plateau transitions in the integer quantum Hall effect (IQHE).

As demonstrated in previous studies of the IQHE\cite{PhysRevLett.61.1297,RevModPhys.67.357}, finite DOS appears on both sides of the extended states (the quantum Hall plateau-transition points), corresponding to the impurity-broadened Landau bands, where the Hall resistance remains quantized. Similarly, just before the topological Anderson transition, a finite-DOS ungapped phase (e.g., at $W\sim10$ eV) also emerges, which is likewise topologically nontrivial. Moreover, under stronger disorder, the impurity-broadened Landau-band region in the IQHE expands, and correspondingly, the ungapped phase of the strongly disordered TACI becomes much wider than that of the weakly disordered T-broken TAI. These parallels indicate that the topological Anderson transition and the quantum Hall transition are closely connected.

In conclusion, we analyze the band structure of ferromagnetic monolayer ${\rm MnBi_4Te_7}$ and construct an effective model to evaluate transport properties under disorder. As disorder increases, the system undergoes a sequence of phase transitions from T-broken QSHE to T-broken TAI, then to TACI, and ultimately to NI. The distinct critical disorder strengths required to destroy the two effective band inversions allow one inversion to survive over a finite parameter window, thereby stabilizing the TACI. The energy window supporting quantized Hall resistance is thus markedly enlarged, enabling higher temperature realizations of the QAHE. Within the effectively band inverted sector, the ZHP regime arises when the disorder-induced in-gap density of states becomes sufficiently large that impurity-mediated scattering substantially degrades the edge channels. The simultaneous emergence of two insulating states, TACI and ZHP insulating state, at different Fermi energies provides complementary experimental signatures supporting our observations.

 The authors would like to thank Hua Jiang, Jiaheng Li and Chuizhen Chen for the helpful discussions.
This work was supported by the National Key R\&D Program of China (Grant No. 2023YFA1607400, 2022YFA1403800), the National Natural Science Foundation of China (Grant No.12274436, 11925408, 11921004), the Science Center of the National Natural Science Foundation of China (Grant No. 12188101). 

\bibliographystyle{unsrt}
\bibliography{ref}
\newpage
\clearpage

       \beginsupplement{}
        \setcounter{section}{0}
        \renewcommand{\thesubsection}{S\arabic{subsection}}
        \renewcommand{\thesubsubsection}{\Alph{subsubsection}}
        \titleformat{\section}[hang]{\bf\centering\Large}{\section}{1.0em}{}[]
        \titleformat{\subsection}[hang]{\bf\large}{\thesubsection}{0.5em}{}[]
        \titleformat{\subsubsection}[hang]{\bf\normalsize}{\thesubsubsection}{0.5em}{}[]

\section*{Supplemental Materials}
\subsection{First-principles calculations}
\begin{figure*}[ht]
\centering
\includegraphics[width=1.0\textwidth]{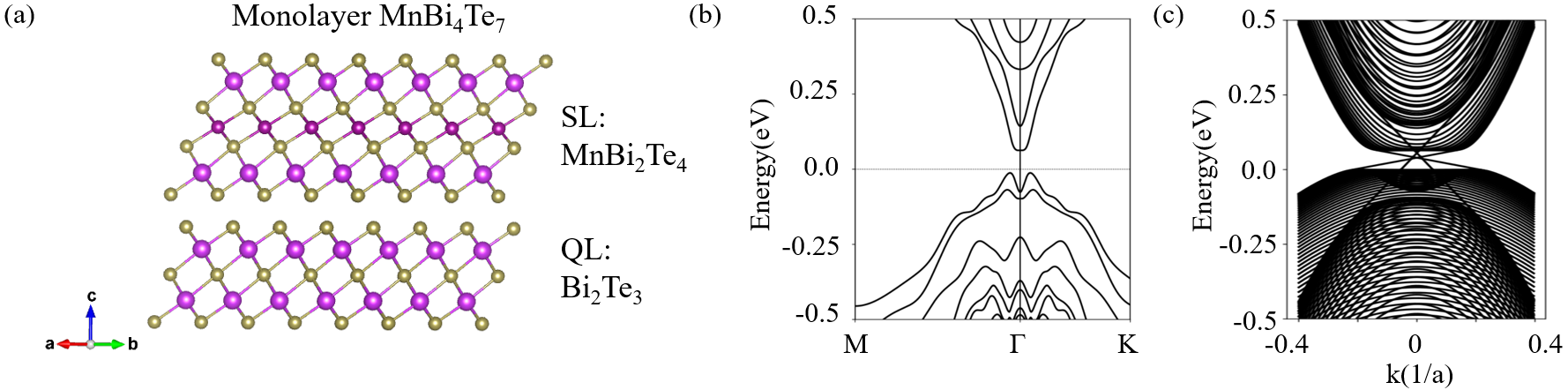}
\caption{First-principles calculations of ferromagnetic(FM) state monolayer ${\rm MnBi_4Te_7}$. (a)Structural schematic of the material structure. (b)Band structures of FM state monolayer ${\rm MnBi_4Te_7}$ by DFT. (c)Band structures of nanoribbon with width $N_y = 200$ by effective k$\cdot$p model.} 
\label{fig:dft}
\end{figure*}

We employed the Vienna ab initio simulation package (VASP) [1] to simulate electronic properties of ferromagnetic monolayer ${\rm MnBi_4Te_7}$, see Fig.\ref{fig:dft} (a), in the framework of density functional theory (DFT)[2,3]. The calculation with the Perdew–Burke–Ernzerhof functional (PBE) [4,5] in the generalized gradient approximation. A cutoff energy of 500 eV and a k-mesh of $8\times8\times1$ are adopted. The calculations are performed with the spin-orbit coupling and U=5 eV for d orbitals of Mn[6,7].

As shown in Fig.\ref{fig:dft} (b). The simulations are in good agreement with prior studies [7]. Near the Fermi level, four $p$ bands are observed: two valence bands and two conduction bands. At the $\Gamma$ point, considering dominant orbital component of the band, these bands are ordered from highest to lowest energy as Te spin-up, Te spin-down, Bi spin-up, and Bi spin-down states, respectively [7]. However, away from the $\Gamma$ point, the Bi orbitals(P2) are positioned at higher energies than the Te orbitals(P1), leading to a band inversion between the Bi and Te orbitals. The spin degeneracy is lifted due to the exchange splitting induced by the magnetic moments from Mn incorporation.

$\mathbf{k} \cdot \mathbf{p}$ Hamiltonian for monolayer $\mathrm{MnBi}_{2n} \mathrm{Te}_{3n+1}$ at the $\Gamma$ point is [8-10]

\begin{gather}
H=H_{\mathrm{N}}(\mathbf{k})+ H_{\mathrm{FM}}(\mathbf{k})
\end{gather}

with the nonmagnetic part $H_{\mathrm{N}}(\mathbf{k})$ and the FM part $ H_{\mathrm{FM}}(\mathbf{k})$ for the correction in the presence of magnetization
$$
H_{\mathrm{N}}=\epsilon_0(\mathbf{k})+\left(\begin{array}{cccc}
M(k) & A_1 k_{-} & 0 &  0\\
A_1 k_{+} & -M(k) & 0 & 0 \\
0 & 0 & M(k) & A_1 k_{-} \\
0 & 0 & A_1 k_{+} & -M(k)
\end{array}\right) ,
$$
and
$$
 H_{\mathrm{FM}}=\left(\begin{array}{cccc}
M_1(k) & A_2 k_{-} & 0 & 0 \\
A_2 k_{+} & -M_2(k) & 0 & 0 \\
0 & 0 & -M_1(k) & -A_2 k_{-} \\
0 & 0 & -A_2 k_{+} &  M_2(k)
\end{array}\right)
$$
in the four-orbital basis of $\left|P 1, \uparrow\right\rangle,\left|P 2, \downarrow\right\rangle,\left|P 1, \downarrow\right\rangle$, and $\left|P 2, \uparrow\right\rangle$. Here $M(k) = M_0 + Bk^{2}, \epsilon_0(k) = C + Dk^2, k_\pm = k_x \pm ik_y, M_{1,2}(k) = M_{1,2} + B_{1,2}k^2,$ and $\epsilon_0(k)$ represents the particle-hole asymmetric term for real material. By fitting to the DFT calculations, the parameters are $M_0 = 0.095 eV, B = -53.49 eV·\text{\AA}^2, M_1 = 0.04 eV, B_1 = -17.44 eV·\text{\AA}^2, M_2 = 0.01eV, B_2 = -8.62 eV·\text{\AA}^2, C = 0.012eV, D = 20.16 eV·\text{\AA}^2, A_1 = 2.82 eV·\text{\AA}, A_2 = -1.8 eV·\text{\AA}$[11]. The effective Hamiltonian is discretized on a square lattice with the lattice constant of ${\rm MnBi_4Te_7}$ \textbf{a} = 4.355 Å. 
 
 Band structures of nanoribbon with width $N_y = 200$ are shown in Fig.\ref{fig:dft} (c), the monolayer $\mathrm{MnBi}_{4} \mathrm{Te}_{7}$ system is a T-broken quantum spin Hall state with two nondegenerate Dirac cones inside the band gap caused by magnetism-induced TRS breaking of degeneracy, and is consistent with the previous calculations [7].

\subsection{The self consistent Born approximation method}

The self energy renormalized by disorder can be self-consistently obtained[12]
\begin{gather}
\Sigma=\frac{W^2}{12} \frac{a^2}{4 \pi^2} \int_{B Z} d^2 \mathbf{k}  \frac{1}{E_f+i \eta-H(\mathbf{k})-\Sigma} 
\end{gather}
where $W$ is the disorder strength,  a randomon-site disorder potential is uniformly distributed in the range $[-\frac{W}{2},\frac{W}{2}]$. $\eta$ is selected to be $10^{-6}$.

After a self-consistent solution $\Sigma$ was obtained, the renormalized parameters are presented

$$\tilde{E}_f=E_f-Re[tr(\Sigma)]$$
$$\tilde{m}_i=m_i+Re\Sigma(i,i)-Re[tr(\Sigma)]$$
where $m_i$ denote the constant on the i-th diagonal element. 

For upper-left block, the region of 
$(\tilde{E}_f-\tilde{m}_1)\times(\tilde{E}_f-\tilde{m}_2)<0$ denote the effecitve band inversion, and 
$(\tilde{E}_f-\tilde{m}_4)\times(\tilde{E}_f-\tilde{m}_3)<0$ for lower-right block. Both values equal to zero give 2 solutions that refer to the $m_1$ to $m_4$ black lines in Fig.1 (d) to denote effective band edges. The effective bulk gaps of different blocks are $\Delta_1 = \tilde{m}_1-\tilde{m}_2$ and $\Delta_2 = \tilde{m}_3-\tilde{m}_4$.

The density of states renormalized by disorder[13] can be obtained by Eq. (4), and for different blocks are $\rho_1$ and $\rho_2$.
In Fig.1 (d), purple lines and blue lines are where $\rho_1>0$ and $\rho_2>0$ to denote the boundaries of the gapped regions and ungapped regions for different blocks. Gray lines and green lines are where $\rho_1=1/\Delta_1$ and $\rho_2=1/\Delta_2$ to denote the boundaries of Anderson transitions. The gray lines show that TACI only exists in electron-doped region while for hole-doped region, it is only a metal-insulator transition. 

The foregoing analysis provides strong evidence that TACI originates from a mismatch between the critical disorder strengths $W_{c1}$ and $W_{c2}$ of the two $2\times 2$ Hamiltonian blocks, at which disorder quenches the corresponding edge states. For $W<\min(W_{c1},W_{c2})$, the system is in the T-broken TAI state. When $W$ enters the intermediate window $\min(W_{c1},W_{c2})<W<\max(W_{c1},W_{c2})$, only one band inversion survives, realizing the TACI state.

\subsection{Edge states projection in the generalized momentum space} 

We compute the disordered energy spectrum in generalized momentum space [14] and analyze the projection of edge-state wavefunctions onto this spectrum. The system size is set to $400\times 15$, we probe the edge at $y=5$ and remove disorder on the outermost boundary to isolate intrinsic edge physics. 

Because ordinary momentum is ill-defined in disordered systems, we construct a quasi-1D superlattice of disordered supercells along $x$, which restores a discrete translational symmetry with boundary phase $\varphi\in[-\pi,\pi]$ of unit cell serving as a generalized momentum. The Hamiltonian is then diagonalized in $(E,\varphi)$, enabling a systematic study of edge evolution with disorder $W$.

\begin{figure*}[ht]
\centering
\includegraphics[width=1.0\textwidth]{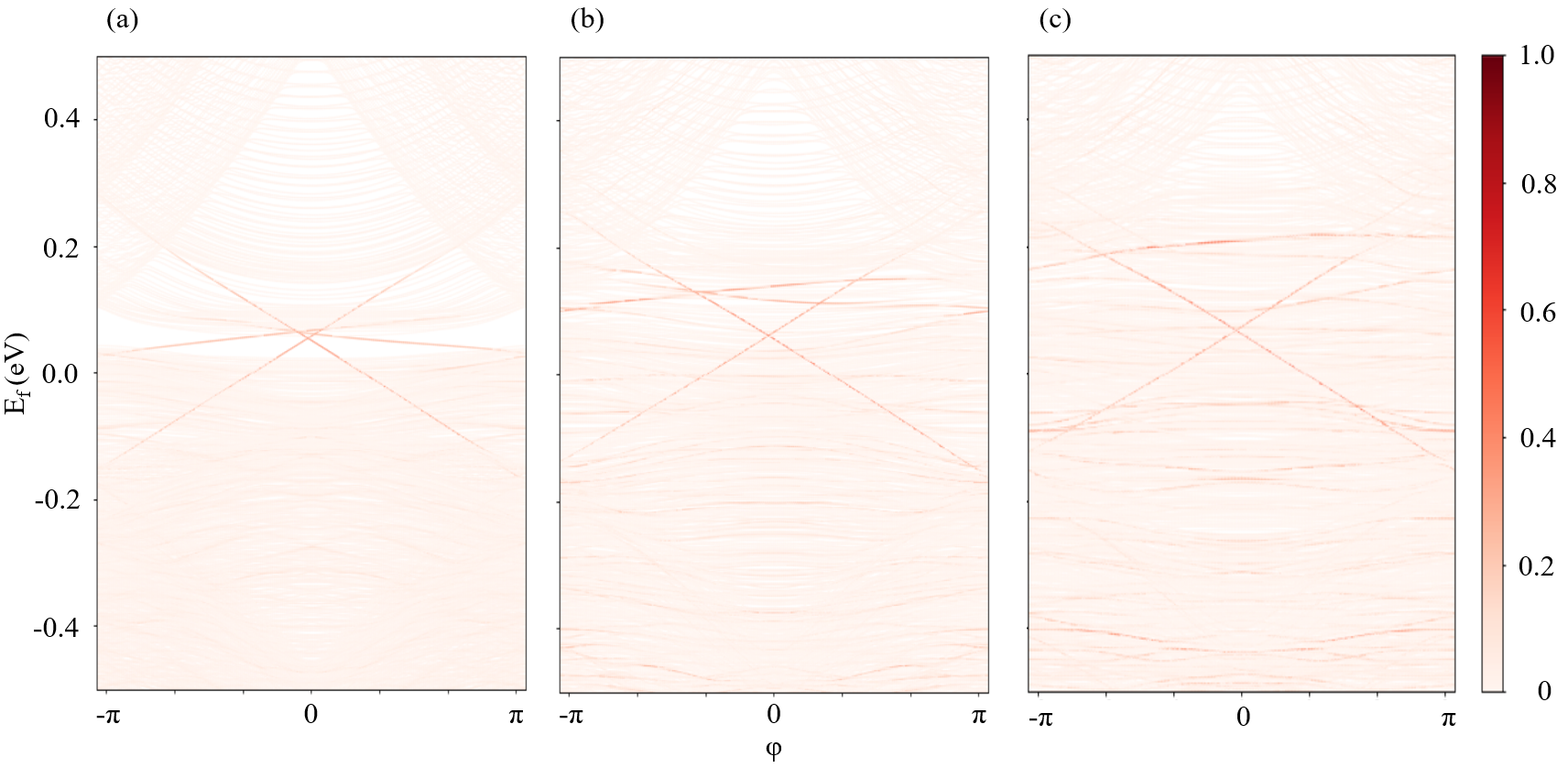}
\caption{Evolutions of edge projection for $W$=0.7, 1.5, 2.5} 
\label{fig:edgepro}
\end{figure*}

Fig.\ref{fig:edgepro} shows the edge-projection evolution for $W=0.7,1.5,2.5$. In Fig.\ref{fig:edgepro} (a), two nondegenerate Dirac cones appear in the gap, which are signatures of the gapped T-broken TAI whose clean counterpart is the T-broken QSHE. In Fig.\ref{fig:edgepro} (b), impurity states progressively fill the gap, signaling a transition from gapped to ungapped behavior and a gradual disappearance of the upper cone. By Fig.\ref{fig:edgepro} (c), the upper cone has vanished entirely, leaving a single Dirac cone characteristic of the TACI. These observations are consistent with the phase diagram.

\subsection{Real space local Chern marker}

To further characterize the topological nature of the disordered finite size samples, we compute the real space local Chern marker (LCM), which provides a gauge invariant real space measure of the Chern number in systems without translational symmetry [15,16]. For a given disorder realization and Fermi energy $E_f$, we first diagonalize the finite Hamiltonian and construct the projector onto the occupied states
\begin{gather}
\hat{P}=\sum_{E_n<E_f}|n\rangle \langle n| ,
\qquad
\hat{Q}=1-\hat{P}.
\end{gather}
The local Chern marker at position $\mathbf{r}$ is defined as [15]
\begin{gather}
C(\mathbf{r})=-2\pi i \langle \mathbf{r}|[\hat{P}\hat{x}\hat{P},\hat{P}\hat{y}\hat{P}]|\mathbf{r}\rangle .
\end{gather}
Equivalently, using $\hat{Q}=1-\hat{P}$, it can be written as
\begin{gather}
C(\mathbf{r})=-4\pi \,\mathrm{Im}\left[\langle \mathbf{r}|\hat{P}\hat{x}\hat{Q}\hat{y}\hat{P}|\mathbf{r}\rangle \right].
\end{gather}
For a lattice system, the marker on site $i$ is
\begin{gather}
C_i=-2\pi i \langle i|[\hat{P}\hat{x}\hat{P},\hat{P}\hat{y}\hat{P}]|i\rangle .
\end{gather}
In a finite sample with open boundaries, $C_i$ exhibits strong edge corrections near the boundary. Therefore, we average the local marker with periodic boundaries.

\begin{figure*}[ht]
\centering
\includegraphics[width=0.8\textwidth]{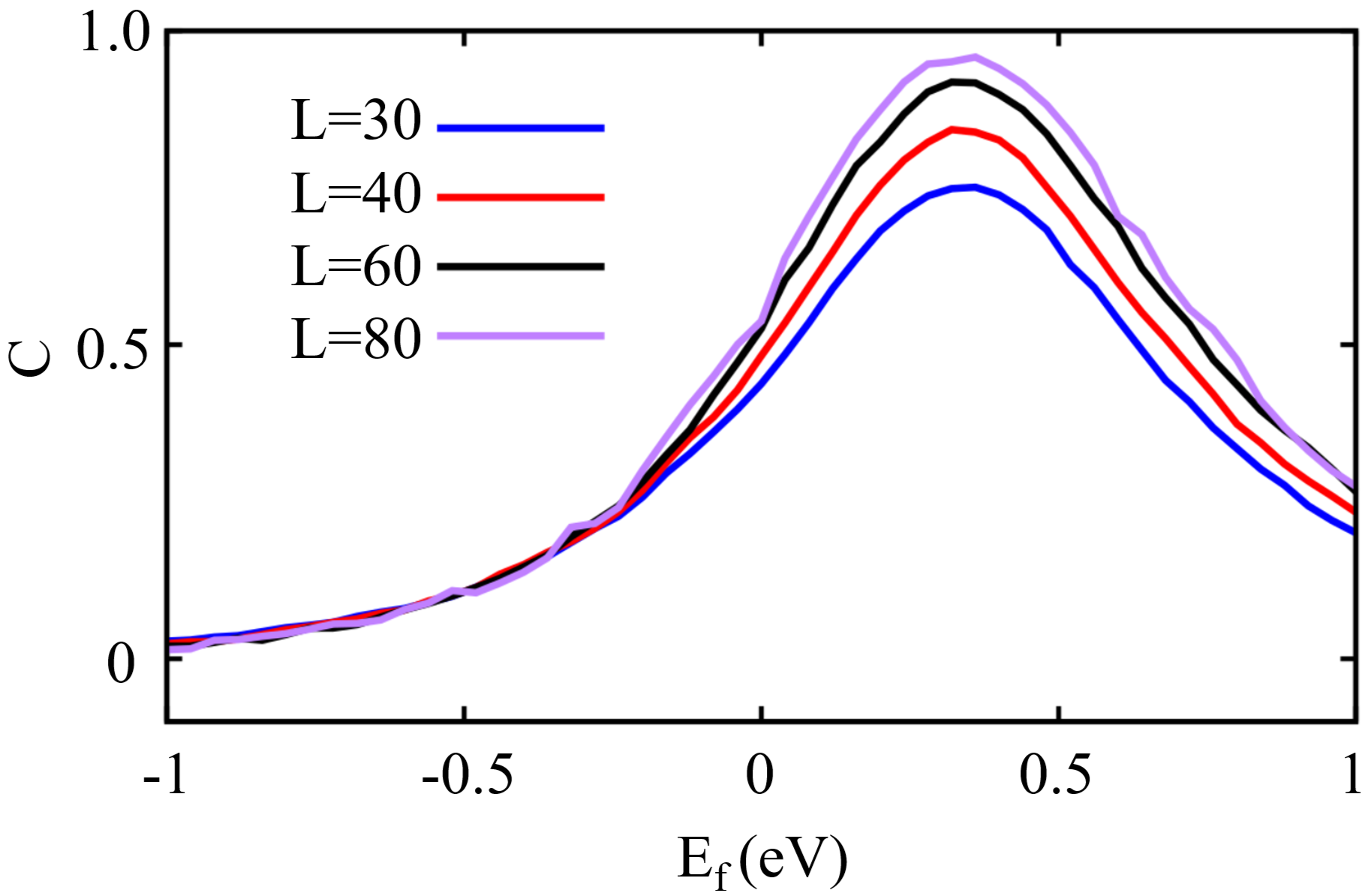}
\caption{ Real space Chern number corresponding to Fig. 2. We present the Chern number as a function of the Fermi energy for different square sample sizes L. For each L, the results are averaged over 100 disorder realizations.} 
\label{fig:chern}
\end{figure*}

For each system size and disorder strength, we further average $\overline{C}_{\mathrm{bulk}}$ over disorder realizations,
\begin{gather}
\langle \overline{C}_{\mathrm{bulk}} \rangle_{\mathrm{dis}}
=
\frac{1}{N_s}\sum_{\alpha=1}^{N_s}\overline{C}_{\mathrm{bulk}}^{(\alpha)} ,
\end{gather}
with $N_s$ the number of disorder samples. In the thermodynamic limit, $\overline{C}_{\mathrm{bulk}}$ converges to the integer Chern number of the occupied states [15]. In the present work, the real space Chern number shown in Fig. \ref{fig:chern} is obtained for the same region as the Fig. 2 in the main text, averaged over 100 realizations and serves as a complementary topological diagnosis to the transport calculations.

[1] G. Kresse and J. Furthmüller, Phys. Rev. B 54, 11169 (1996).

[2] P. Hohenberg and W. Kohn, Phys. Rev. 136, B864 (1964). 

[3] W. Kohn and L. J. Sham, Phys. Rev. 140, A1133 (1965).

[4] J. P. Perdew, K. Burke, and M. Ernzerhof, Phys. Rev. Lett. 78, 1396 (1997). 

[5] J. P. Perdew, K. Burke, and M. Ernzerhof, Phys. Rev. Lett. 77, 3865 (1996).

[6] M. M. Otrokov et al., 2D Mater. 4, 025082 (2017).

[7] H. Sun, B. Xia, Z. Chen, Y. Zhang, P. Liu, Q. Yao, H. Tang, Y. Zhao, H. Xu, and Q. Liu, Phys. Rev. Lett. 123, 096401 (2019).

[8] Dongqin Zhang et al., Topological axion states in the magnetic insulator mnbi2te4 with the quantized magnetoelectric effect. Phys. Rev. Lett. 122:206401 (2019).

[9] Chui-Zhen Chen et al., Evolution of berry curvature and reentrant quantum anomalous hall effect in an intrinsic magnetic topological insulator. Science China Physics, Mechanics \& Astronomy, 64(12):127211 (2021).

[10]Biao Lian et al., Flat chern band from twisted bilayer mnbi2te4. Phys. Rev. Lett., 124:126402 (2020).

[11]Anqi Wang et al., Observation of topological anderson chern insulator phase in mnbi4te7 monolayer, 2025. arXiv:2501.04354 

[12] C. W. Groth et al. , Phys. Rev. Lett. 103, 196805 (2009).

[13] DinhDuy Vu and Sankar Das Sarma. Phys. Rev. B, 106:134201 (2022).

[14] Liang Chen et al., New Journal of Physics, 14(4):043028 (2012).

[15] R. Bianco and R. Resta, Phys. Rev. B 84, 241106(R) (2011).

[16] R. Resta, Eur. Phys. J. B 79, 121 (2011).

\clearpage

\end{document}